\documentclass[aps, prb, twocolumn, superscriptaddress, 10pt, floatfix]{revtex4-1}
\usepackage{graphicx}
\usepackage[hidelinks]{hyperref}

\begin{document}

\title{Andreev molecule in parallel InAs nanowires}

\author{Oliv\'er K\"urt\"ossy}
\affiliation{Department of Physics, Budapest University of Technology and Economics and Nanoelectronics 'Momentum' Research Group of the Hungarian Academy of Sciences, Budafoki \'ut 8, 1111 Budapest, Hungary}

\author{Zolt\'an Scher\"ubl}
\affiliation{Department of Physics, Budapest University of Technology and Economics and Nanoelectronics 'Momentum' Research Group of the Hungarian Academy of Sciences, Budafoki \'ut 8, 1111 Budapest, Hungary}

\affiliation{Univ. Grenoble Alpes, CEA, Grenoble INP, IRIG, PHELIQS, 38000 Grenoble, France}

\author{Gerg\H o F\"ul\"op}
\affiliation{Department of Physics, Budapest University of Technology and Economics and Nanoelectronics 'Momentum' Research Group of the Hungarian Academy of Sciences, Budafoki \'ut 8, 1111 Budapest, Hungary}

\author{Istv\'an Endre Luk\'acs}
\affiliation{Center for Energy Research, Institute of Technical Physics and Material Science, Konkoly-Thege Mikl\'os \'ut 29-33., H-1121, Budapest, Hungary}

\author{Thomas Kanne}
\affiliation{Center  for  Quantum  Devices,  Niels  Bohr  Institute,University  of  Copenhagen,  2100  Copenhagen,  Denmark}

\author{Jesper Nyg{\aa}rd}
\affiliation{Center  for  Quantum  Devices,  Niels  Bohr  Institute,University  of  Copenhagen,  2100  Copenhagen,  Denmark}

\author{P\'eter Makk}
\email{peter.makk@mail.bme.hu}
\affiliation{Department of Physics, Budapest University of Technology and Economics and Nanoelectronics 'Momentum' Research Group of the Hungarian Academy of Sciences, Budafoki \'ut 8, 1111 Budapest, Hungary}

\author{Szabolcs Csonka}
\email{szabolcs.csonka@mono.eik.bme.hu}
\affiliation{Department of Physics, Budapest University of Technology and Economics and Nanoelectronics 'Momentum' Research Group of the Hungarian Academy of Sciences, Budafoki \'ut 8, 1111 Budapest, Hungary}

\begin{abstract}
Coupling individual atoms via tunneling fundamentally changes the state of matter: electrons bound to atomic cores become delocalized resulting in a change from an insulating to a metallic state, as it is well known from the canonical example of solids. A chain of atoms could lead to more exotic states if the tunneling takes place via the superconducting vacuum and can induce topologically protected excitations like Majorana or parafermions. Toward the realization of such artificial chains, coupling a single atom to the superconducting vacuum is well studied, but the hybridization of two sites via the superconductor was not yet reported. The peculiar vacuum of the BCS condensate opens the way to annihilate or generate two electrons from the bulk resulting in a so-called Andreev molecular state. By employing parallel nanowires with an Al superconductor shell, two artificial atoms were created at a minimal distance with an epitaxial superconducting link between. Hybridization via the BCS vacuum was observed between the two artificial atoms for the first time, as a demonstration of an Andreev molecular state.
\end{abstract}

\date{\today}



\maketitle

\begin{figure*}[htp]
\includegraphics[width=1\textwidth]{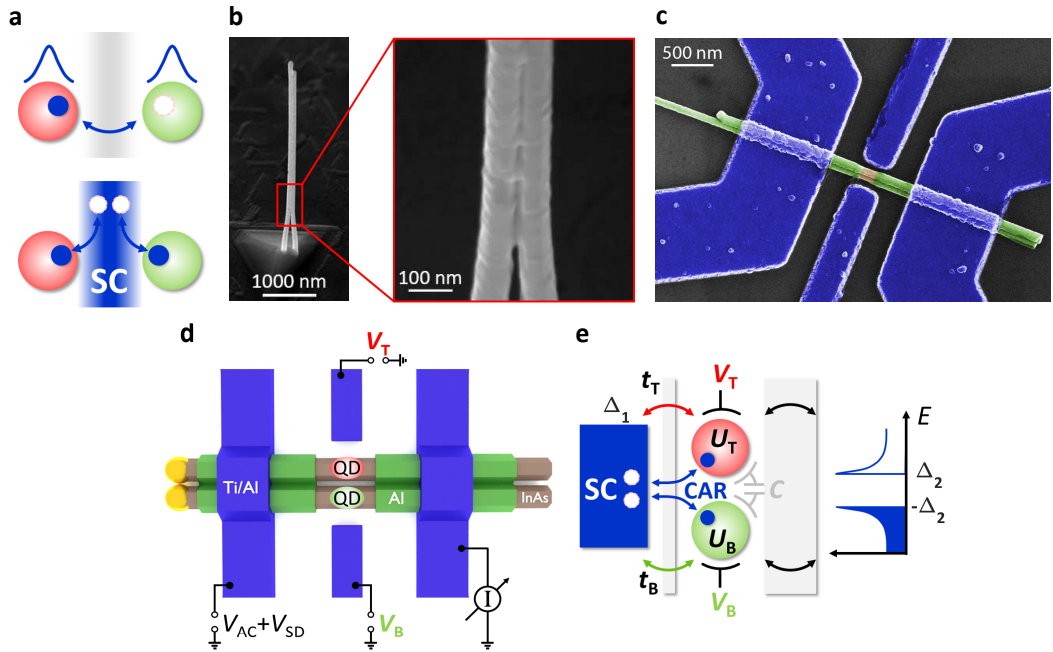}
\caption{\label{device_outline} \textbf{Device outline.} \textbf{a} General concept of a molecular state formed by tunneling between two atomic sites (top). In our case, the interaction between the quantum dots (QDs) is mediated by a superconductor (SC, bottom), where two electrons can be created from the SC vacuum. \textbf{b} High-resolution scanning electron micrograph (SEM) of the as-grown parallel wires. The epitaxial Al connects the two InAs nanowires. \textbf{c} False colored SEM of device B. Brown: InAs, green: epitaxial Al, blue: evaporated Ti/Al. \textbf{d} Schematic illustration of the device and the measurement setup. Differential conductance $G=I_\mathrm{AC}/V_\mathrm{AC}$ was recorded in two-terminal measurements. The QDs in the top and bottom wires were tuned by plunger gate voltages $V_\mathrm{T}$ and $V_\mathrm{B}$, respectively. \textbf{e} Sketch of the setup used for modeling the system with tunnel coupling $t_\mathrm{T}$ ($t_\mathrm{B}$), charging energy $U_\mathrm{T}$ ($U_\mathrm{B}$) and on-site energy $\varepsilon_\mathrm{T}$ ($\varepsilon_\mathrm{B}$) controlled by $V_\mathrm{T}$ ($V_\mathrm{B}$) belonging to the top (bottom) QD. Grey rectangles illustrate the tunnel barriers. While the left electrode with a gap of $\Delta_1$ was strongly coupled to the QDs, the right one was weakly coupled, thus the latter served as a BCS probe with a gap of $\Delta_2$. Interdot capacitance $C$ was also considered.}
\end{figure*}

Based on Bardeen-Cooper-Schrieffer (BCS) mean-field theory\cite{PhysRev.108.1175}, the superconducting vacuum only allows to add individual electrons with energy above the superconducting gap, however, it serves as a free source and drain of electron pairs, known as Cooper pairs (see Fig. \ref{device_outline}\textbf{a}). The interplay between an artificial atom, namely a quantum dot (QD), and the BCS vacuum were studied intensively, contributing to the formation of a sub-gap excitation, a so-called Yu-Shiba-Rusinov (YSR) state (or called Andreev Bound states in other limits)\cite{yu1965bound,shiba1968classical,rusinov1969theory,balatsky2006impurity,buitelaar2002quantum,sand2007kondo,eichler2007even,grove2009superconductivity,pillet2010andreev,lee2014spin,jellinggaard2016tuning,scherubl2020large,prada2020andreev}. These excitations are shared between the QD and the superconductor (SC)\cite{scherubl2020large}. They naturally raise the prospect of hybridized states: two bound states that are separated by the SC, and which we call an Andreev-molecule\cite{scherubl2019transport}. Coupling two QDs to a joint SC is also a basic building block of a Cooper pair splitter (CPS)\cite{recher2001andreev}, where the QDs are attached to two normal leads allowing to create spatially separated entangled electron pairs\cite{hofstetter2009cooper,herrmann2010carbon,hofstetter2011finite,schindele2012near,das2012high,deacon2015cooper} via crossed Andreev reflection (CAR)\cite{byers1995probing,deutscher2000coupling,lesovik2001electronic}. While a CPS favors weak SC-QD couplings, an Andreev molecule requires the opposite limit. Several theoretical works investigated the Andreev molecular state \cite{eldridge2010superconducting,sau2012realizing,trocha2015spin,wrzesniewski2017current,scherubl2019transport,pillet2020scattering}, e.g. as a minimal model for Majorana chains \cite{leijnse2012parity}, nevertheless, the realization of hybridized parallel YSR states via a common SC has not been reported yet. This is due to the fact that creation and detection impose a set of challenging constraints on the device: the QDs must be strongly coupled to the SC, their distance should be minimized while preventing direct tunneling between them\cite{fulop2014local,fulop2015magnetic}, in addition, individual tunability of the atomic levels is also desired. We construct our artificial atoms in a novel superconducting hybrid nanostructure, where double InAs nanowires are grown in close vicinity and connected by an epitaxial SC Al shell, allowing us to fulfill the requirements outlined above\cite{kanne2021double,vekrispreprint2021,vekrispreprintLP2021}.

InAs nanowires with epitaxial Al shell became a highly attractive nanostructure for superconducting hybrid quantum devices, like Andreev-qubits\cite{zazunov2003andreev,janvier2015coherent,hays2018direct,tosi2019spin}, Gatemons\cite{nakamura1999coherent,larsen2015semiconductor}, or Majorana devices\cite{kitaev2001unpaired,lutchyn2010majorana,oreg2010helical,mourik2012signatures,das2012zero,deng2016majorana,albrecht2016exponential}, due to the perfect SC-semiconductor interface providing a strong proximity effect\cite{krogstrup2015epitaxy,chang2015hard}. Our device is built on a versatile epitaxial superconductor hybrid, where two separated adjacent wires are covered by a common epitaxial Al shell, allowing to define QDs in close proximity with the absence of direct coupling (see Fig. \ref{device_outline}\textbf{b}). Recent works have already reported the Cooper pair splitting signals\cite{baba2018cooper} and non-local pair tunnelings\cite{ueda2019dominant} in individual nanowires placed parallel close to each other manually with a micromanipulator. Furthermore, Andreev bound states were also coupled by direct tunneling between two serial QDs\cite{su2017andreev} and in a dimer \cite{kezilebieke2018coupled}. However, none of them has realized the strong hybridization of artificial atoms via a SC needed for the formation of the Andreev molecular state, the elementary building block of a Majorana-chain. In this paper, we report the signature of the Andreev molecular state, for the first time in parallel InAs nanowires. We discuss first the case of non-interacting YSR states and then compare it to the strongly interacting system involving the hybridization via SC and Coulomb repulsion, both experimentally and theoretically.

\textbf{Device outline.} The specific system studied here is illustrated in Fig. \ref{device_outline}\textbf{c-e}. A parallel double QD was formed in a pair of InAs nanowires merged by epitaxial full-shell Al\cite{kanne2021double}, which was etched away on a $\sim$250 nm long segment. Two common superconducting (Ti/Al) electrodes were attached to epitaxial Al on the nanowires forming parallel SC-QD-SC junctions in the two wires. Low-temperature electronic transport measurements were carried out at a base temperature of 40 mK (for details see Methods). In two-terminal sub-gap spectroscopy, the differential conductance $G=I_\mathrm{AC}/V_\mathrm{AC}$ was measured with the tuning of the QDs by individual plunger gates, as depicted in Fig. \ref{device_outline}\textbf{d} ($V_\mathrm{T}$ corresponds to the top, $V_\mathrm{B}$ to the bottom gate voltage). The source terminal biased with $V_\mathrm{SD}$ was found to be coupled strongly to the QDs, whereas the other one worked as a SC tunnel probe leading to a SC-QD-I-SC junction, where I stands for insulator. Two different devices are presented in this paper; one did not show strong coupling between nanowires, and thus, serves as a reference junction (device A), whereas for the other a strong hybridization of the QDs and signatures of the molecular states was observed (device B).
 
\begin{figure*}[htp]
\includegraphics[width=1\textwidth]{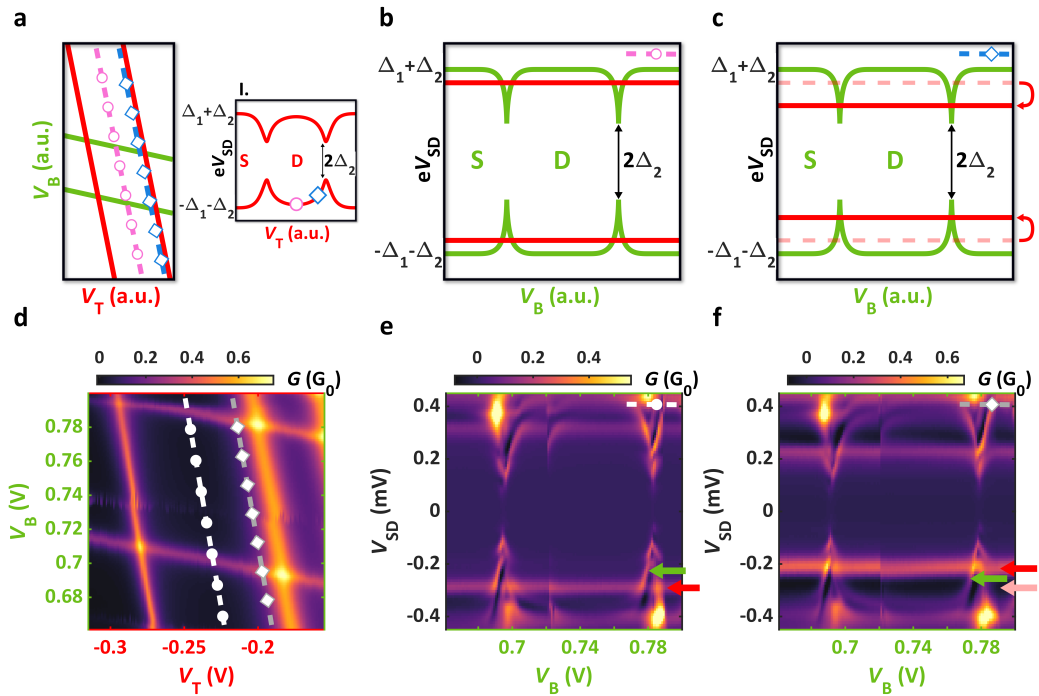}
\caption{\label{non_int_Shiba} \textbf{Non-interacting parallel YSR states (device A).} \textbf{a} Schematic illustration about the stability map of parallel QDs with joint electrodes. Resonances of the top and bottom QDs are depicted with red and green, respectively. The lever arms refer to a finite cross capacitance of each gate to the opposite QDs. Inset \textbf{I.} depicts the YSR spectrum residing in the top QD ("S" and "D" refer to singlet and doublet ground states, respectively). $\Delta_1$ and $\Delta_2$ are the superconducting gaps of the strongly coupled electrode and the SC probe. The pink circle and the blue diamond indicate the energy of YSR\textsubscript{T} along the cuts taken in the stability map. \textbf{b} Expected excitation spectrum of YSR states along the pink line shown in panel \textbf{a}. Whereas the local YSR\textsubscript{B} state (green) evolves along the cut since it is sensitive to its own gate ($V_\mathrm{B}$), the non-local YSR\textsubscript{T} (red) stays on constant energy as the slice is parallel to the red resonances. \textbf{c} Bound state spectrum along the blue line in panel \textbf{a}. The excitation of the YSR\textsubscript{T} state moved to lower energy compared to the one in panel \textbf{b} (the original energy is depicted with pink dashed lines) as the charge degeneracy of the top QD was approached (see the blue diamond in inset \textbf{I.} of panel \textbf{a}). \textbf{d} Measured conductance as a function of gate voltages for device A in the normal state. \textbf{e} Finite-bias spectroscopy measurement along the white dashed line depicted in panel \textbf{d} (superconducting state). Panel \textbf{b} illustrates well the experimental findings. Green and red arrows mark the excitations of YSR\textsubscript{B} and YSR\textsubscript{T}, respectively. \textbf{f} Finite-bias spectroscopy measurement along the gray dashed line with a diamond in panel \textbf{d}, closer to the resonance of the top QD matching to panel \textbf{c}. The pink arrow indicates the position of the non-local signals in panel \textbf{e}.}
\end{figure*}

In QDs coupled strongly to SC electrodes, sub-gap states are formed by hybridization between a QD level and the SC electrode\cite{yu1965bound,shiba1968classical,rusinov1969theory,jellinggaard2016tuning}. For Coulomb repulsion energies larger than the superconducting gap (YSR limit), the relevant number of electrons in the QD and quasi-particles in the SC is restricted to 0 or 1. If the total particle number is odd (the QD is singly occupied and the SC is empty or vice versa), the system is in the doublet ground state, while if it is even (empty state or a pair of electron and quasi-particle in the system) a singlet state is formed. A finite tunnel coupling between the QD and the SC hybridize two states with the same parity, and ground state transitions can be induced from the doublet to the singlet state by changing the on-site energy of the QD (for details see Ref. \onlinecite{prada2020andreev}). In transport experiments, the excitation energies are probed via finite-bias spectroscopy leading to "eye-shaped" excitations as a function of a plunger gate voltage\cite{lee2014spin,jellinggaard2016tuning}, similarly to the one sketched in inset \textbf{I.} of Fig. \ref{non_int_Shiba}\textbf{a}. 

In the following, we review the spectrum of the parallel double QD structure in three steps: i) two independent, non-interacting YSR states, ii) adding interdot Coulomb repulsion, and iii) including the superconducting coupling. We label the QDs and their features as top (T) and bottom (B) ones, marked with red and green as in Fig. \ref{device_outline}, respectively, supposing only a single YSR state residing in each QD (YSR\textsubscript{T} and YSR\textsubscript{B}). 

\textbf{Non-interacting YSR states.} The parallel YSR states in uncoupled wires are discussed in Fig. \ref{non_int_Shiba} (top row expectations, bottom row measurements). Panel \textbf{a} illustrates the zero-bias conductance of the two wires as a function of the two plunger gate voltages in the normal state. Here the interdot capacitance is negligible, however, there is a finite cross-capacitance between the top (bottom) plunger gate and the bottom (top) QD, resulting in the tilted lines in the phase diagram. Panels \textbf{b} and \textbf{c} illustrate the finite-bias spectrum along the pink and blue dashed line in panel \textbf{a} parallel to the top QD resonances as a function of $V_\mathrm{B}$. Asymmetric coupling of the wires, $t_\mathrm{T}>t_\mathrm{B}$ and different superconducting gaps of $\Delta_1$ and $\Delta_2$ for the strongly coupled electrode and the SC probe are considered, respectively, to reproduce the experimentally observed features. The spectroscopy yield the sum of an "eye-shaped" excitation (green) typical for YSR systems and an excitation line at constant energy (red). The green YSR patterns belong to the bottom QD (YSR\textsubscript{B}, called local signal), while the red ones can be identified as YSR\textsubscript{T} (called non-local signal) since the on-site energy of the top QD is kept constant due to the parallel slicing in both panels. The excitations do not touch at zero $V_\mathrm{SD}$, but stay always at finite energy originating from the SC tunnel probe, which introduces a $\pm \Delta_2$ gap in the excitation spectrum. These minima correspond to the ground state transitions of the local signal addressed also in the figure ("D" and "S" stand for doublet and singlet ground states, respectively). Depending on the position of the slice, the energy of the constant line can vary between $\Delta_2$ and $\Delta_1+\Delta_2$. Obviously, the non-local YSR state can occupy the lowest energy $\Delta_2$ when the corresponding (top) QD is close to (panel \textbf{c}, blue line in panel \textbf{a}), while moving deeper in the blockade brings its energy towards the gap edge $\Delta_1+\Delta_2$ regardless of the parity of the ground state. The movement of the non-local signal while approaching a resonance is indicated with red arrows in panel \textbf{c}. For clarity, inset \textbf{I.} in panel \textbf{a} depicts YSR\textsubscript{T} as the function of its own plunger gate ($V_\mathrm{T}$), in which the markers assign the actual excitation energies considered in panels \textbf{b} and \textbf{c}.

\begin{figure*}[htp]
\includegraphics[width=1\textwidth]{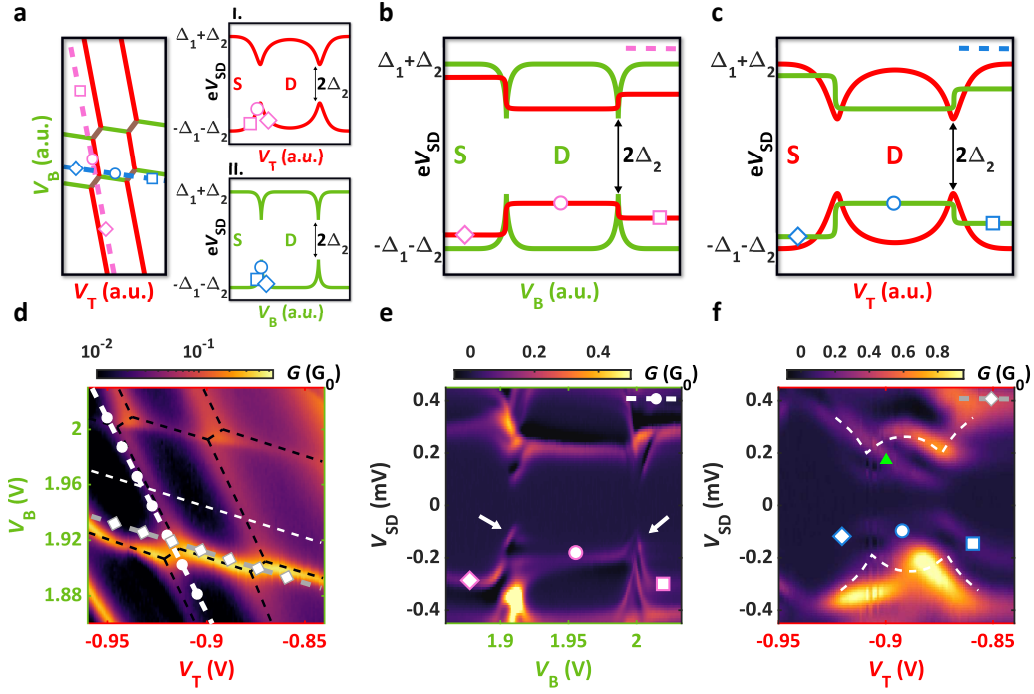}
\caption{\label{int_Shiba} \textbf{Capacitively coupled YSR states (device B).} \textbf{a} Stability map of parallel QDs with strong interdot capacitance. Despite the pink and blue cuts being parallel to the resonances of the top and bottom QDs, respectively, the excitation energies of the non-local YSR states shift while crossing the triple points. Markers in the insets indicate the non-local YSR state energies along the cuts. \textbf{b} Predicted spectrum along the pink line (tuning the bottom QD) from panel \textbf{a} by considering the Coulomb interaction. In this case, steps in the energy of YSR\textsubscript{T} are expected. \textbf{c} Similar spectrum to the one in panel \textbf{b}, but along the blue line in panel \textbf{a} resulting in the tuning of the top QD. "Step-like" features (green curve) in the YSR\textsubscript{B} develop again. \textbf{d} Normal state conductance as a function of the plunger gates on device B. Black dashed line illustrates the honey-comb pattern attributed to the parallel QD system. \textbf{e} Bias spectroscopy measurement along the dashed line with circles depicted in panel \textbf{d}. Whereas most features show resemblance to panel \textbf{b}, anti-crossings and bends towards zero energy occur at the charge degeneracy points marked by the white arrows. \textbf{f} Measured spectrum along the other dashed line with diamonds from panel \textbf{d}. The white dashed line shows the local "eye-shaped" YSR doublet when it is recorded off-resonance with its trace depicted in panel \textbf{d}. Additional excitation lines, such as the one marked by the green triangle, also appear. In comparison to panel \textbf{c}, it can be seen that the measurements can not be described by the simple capacitive model.}
\end{figure*}

The measurements of device A follow well our basic predictions outlined above. The stability map in the normal state, which was recorded by applying 250 mT out-of-plane magnetic field, is shown in Fig. \ref{non_int_Shiba}\textbf{d}. Two different bias cuts parallel to the top QD resonance in panels \textbf{e} and \textbf{f} reveal the movement of the non-local YSR state (red arrow) while the development of the local YSR state (green arrow) being intact. We note that similar behavior of the bound states was observed along cuts parallel to the bottom QD resonances (for details see Supplementary Note 1). Based on the two dominant lever arms in the stability sweep and the fact that different YSR states were captured by tuning either $V_\mathrm{T}$ or $V_\mathrm{B}$ confirmed the model of having YSR states in both QDs. 

\textbf{Interdot Coulomb repulsion.} The question arises of how the spectrum is modified compared to the non-interacting case if there is significant interdot capacitance (see $C$ in Fig. \ref{device_outline}\textbf{e}). The normal state stability map turns into the so-called honey-comb pattern (see Fig. \ref{int_Shiba}\textbf{a}) well known for double QDs\cite{van2002electron}. Hence, slicing parallel to any resonances along a straight line in the gate map gives no longer a constant-energy non-local YSR state, but a charge state-dependent one. For example, along the pink dashed line, the non-local YSR state develops according to the symbols in inset \textbf{I.}, where YSR\textsubscript{T} is depicted as a function of its own plunger gate, $V_\mathrm{T}$. For small $V_\mathrm{B}$ values, the line cut is off-resonance and YSR\textsubscript{T} is in the doublet ground state (diamond symbol). By increasing $V_\mathrm{B}$ the bottom QD is brought to resonance effectively gating the top QD and shifting it closer to its resonance, which lowers the energy of YSR\textsubscript{T} (circle symbol). Going through another resonance of the bottom QD (by further increasing $V_\mathrm{B}$) displaces the top QD resonance again and its YSR state ends up in the singlet ground state (square symbol). These jumps of the non-local signal at the charge degeneracy points, where the on-site energy of the QD changes abruptly, imply "step-like" excitations in total (see the red lines in panel \textbf{b}). Analogously, a similar spectrum (shown in panel \textbf{c}) is obtained along the blue line in panel \textbf{a} with the symbols in inset \textbf{II.}.

Whereas the uncoupled YSR states described well device A, they clearly can not match the measurements on device B, shown in the bottom row of Fig. \ref{int_Shiba}. Therefore we now compare them to the simple case of having capacitive coupling between the two QDs. Fig. \ref{int_Shiba}\textbf{d} shows the measured normal state map providing qualitatively the same honey-comb structure (illustrated with black dashed lines) as the one in panel \textbf{a}. Sub-gap spectroscopy was performed along the white and gray dashed lines. Similarly to device A, in the measurements of device B the conductance of YSR\textsubscript{T} (red) was found to be larger than YSR\textsubscript{B} (green) fulfilling the assumption of $t_\mathrm{T}>t_\mathrm{B}$ already mentioned. In panel \textbf{e}, the spectrum is presented as a function of $V_\mathrm{B}$ exhibiting similarities to panel \textbf{b}. The local YSR\textsubscript{B} state is mostly bound to the gap edge and develops rapidly at the ground state transitions matching the green curve in panel \textbf{b}. The non-local YSR\textsubscript{T} state (marked by the pink symbols at negative bias) also provides "step-like" features in accordance with the red curve in panel \textbf{b} (marked by the pink symbols). Nonetheless, clear discrepancies emerge close to the charge degeneracy points ($V_\mathrm{B}=1.91\,$V and $V_\mathrm{B}=2\,$V). As the local and non-local excitations approach each other, they anti-cross, and the non-local one bends towards zero energy (indicated by white arrows) suggesting the hybridization of the states, which is unexpected in a simple capacitive picture. The difference between the theoretical spectrum and the measured data is more prominent for cuts along the other gate direction as the comparison of panels \textbf{c} and \textbf{f} shows. Assuming only capacitive coupling between QDs (panel \textbf{c}), YSR\textsubscript{T} is expected to take the red, "eye-shaped" curve as a function of $V_\mathrm{T}$. Such excitation is measured when YSR\textsubscript{B} is far off-resonance (see the white dashed line in panel \textbf{f} and its trace in panel \textbf{d}, and also Supplementary Note 2). However, the spectrum captured close to the resonance of YSR\textsubscript{B} (along the gray dashed line in panel \textbf{d}) strongly deviates from the expectation of simple capacitive coupling as the comparison of panel \textbf{f} and \textbf{c} demonstrates: (i) the "eye-shaped" YSR\textsubscript{T} resonance is completely distorted in the measurements. Moreover, (ii) the expected horizontal non-local signal does not stay flat in the doublet region of YSR\textsubscript{T} (indicated by the blue circle between $V_\mathrm{T}=-0.92\,$V and $V_\mathrm{T}=-0.87\,$V), rather follows the curvature of the local signal. Though well-pronounced anti-crossings are absent, (iii) extra dispersive lines (one example is marked by the green triangle) appear between the local and non-local signal in energy. (iv) It is also remarkable that the measured spectrum is asymmetric for the sign of the bias. The highlighted features and the unusual evolution of the non-local signals were quite robust along any cuts taken in the vicinity of the charge degeneracies (for further data see Supplementary Note 2 and 3). All these observations suggest that the YSR states in the QDs interact with each other via the SC electrode forming an Andreev molecule. To strengthen this finding, we present the results of a minimal model of a fully interacting two-dot system, which reproduces the experimental signatures listed above.

\begin{figure}[htp]
\includegraphics[width=0.5\textwidth]{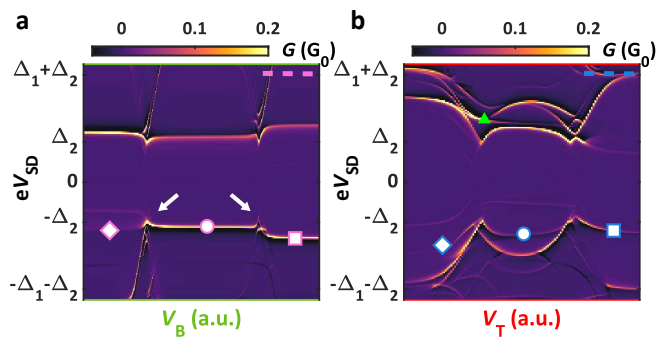}
\caption{\label{simulations} \textbf{Simulations of the hybridized YSR states.} \textbf{a} Numeric simulation of Fig. \ref{int_Shiba}\textbf{b} with allowing hybridization via the SC. The superconductivity induces bendings and anti-crossings in both local and non-local YSR states at the charge degeneracy points. The excitations of the local, weakly coupled YSR\textsubscript{B} state are also enhanced. \textbf{b} Similar simulation, but following the gate settings of \ref{int_Shiba}\textbf{c}. Besides the non-local YSR\textsubscript{B} state being enhanced again, it also develops distinctly from the capacitive model and leads to distortions in the local signal. As one can see, the excitation lines are multiplied and not even symmetric in bias. The non-local signal is not "step-like", but follows the curvature of the local one. In the model we have used charging energies $U_\mathrm{T}=1.2\,$meV and $U_\mathrm{B}=2.2\,$meV, off-site repulsion energy (proportional to the interdot capacitance) $U_\mathrm{C}=0.1\,$meV, and superconducting gaps of $\Delta_1 = 200 \ \mu$V and $\Delta_2 = 120 \ \mu$V, which were extracted from the measurements. Tunnel amplitudes were estimated as $t_\mathrm{T}=0.15\,$meV and $t_\mathrm{B}=0.05\,$meV.}
\end{figure}

\textbf{Superconducting coupling.} As seen from the comparison of the sketches and measurements in Fig. \ref{int_Shiba}, the theory requires to go beyond the interdot Coulomb interaction to interpret the experimental data. To proceed, we applied a model that involved SC-induced hybridization between the QDs. The dots are modeled by capacitively interacting single sites (see Fig. \ref{device_outline}\textbf{e}) coupled to the SC with tunnel amplitudes $t_\mathrm{T}$ and $t_\mathrm{B}$. The SC is handled in the zero bandwidth approximation\cite{affleck2000andreev,probst2016signatures} and the coupling to the right BCS probe is treated perturbatively. The excitation spectra are derived with exact diagonalization of the Fock-space Hamiltonians, from which the transport is numerically calculated by solving the classical master equation using Fermi's golden rule (for more details see Supplementary Note 4). The simulated spectra taken along the pink and blue lines depicted in Fig. \ref{int_Shiba}\textbf{a} are shown in Fig. \ref{simulations}.

In the purely capacitively interacting case, the simulations give back the spectra illustrated in Fig. \ref{int_Shiba} panels \textbf{b} and \textbf{c} (for details see Supplementary Note 2). In turn, when the two QDs are hybridized via the SC, as shown in Fig. \ref{simulations}, the spectra are significantly different from the non-interacting ones and they are in good agreement with the measurements introduced in Fig. \ref{int_Shiba}\textbf{e-f}. The anti-crossings and the bendings of YSR\textsubscript{T} observed particularly in Fig. \ref{int_Shiba}\textbf{e} are recovered in the numerical results (see white arrows in Fig. \ref{simulations}\textbf{a}). We emphasize that, contrary to simple expectations, anti-crossings do not necessarily appear in the presence of strong hybridization as seen in panel \textbf{b}. This finding is also in accordance with the experiment in Fig. \ref{int_Shiba}\textbf{f}. The dispersive evolution of the signals in panel \textbf{b} is also pronounced in agreement to the data in Fig. \ref{int_Shiba}\textbf{f}, where the lower energy, non-local YSR\textsubscript{B} state (indicated with the blue markers again) establishes a similar shape to the local one. It is notable that the conductance of the YSR\textsubscript{B}, whose coupling is three times weaker than YSR\textsubscript{T} in the model, is greatly enhanced and reaches $\sim$80\% of the strongly coupled YSR\textsubscript{T} near the charge degeneracy points, which was also captured in the measurements. Additional excitation lines (see the green triangle) being asymmetric in bias are also present in the simulations (Fig. \ref{simulations}\textbf{b}) matching the experimental data (Fig. \ref{int_Shiba}\textbf{f}). Despite the model being simplified, most characteristic features of the measurements were captured by adding hybridization via the SC, which supports our interpretation of the observation of the Andreev molecule. 

The two investigated devices (A and B) showed very different behaviors. Whereas for A no hybridization was observed, sample B exhibited signatures of the Andreev molecule. Careful SEM analysis revealed an important structural difference; for sample B the two InAs nanowires were merged by the epitaxial Al, while for device A, the wires have separated and became only connected by the ex-situ evaporated contacts (blue in Fig. \ref{device_outline}\textbf{a}, see Supplementary Note 1). These Ti/Al contacts were established$•$ $\sim$400 nm away from the QDs, which could explain the absence of the SC-induced hybridization in device A\cite{recher2001andreev,anselmetti2019end}.

In summary, we have found strong interactions between parallel YSR states realized in double InAs nanowires connected by an epitaxial Al shell. The small geometrical distance between the QDs resulted in capacitive coupling, while the shared epitaxial Al source contact enabled hybridization via the SC vacuum. The latter one allowed the emergence of an Andreev molecular state, whose transport signatures were measured for the first time. The detected spectroscopic features were reproduced by our numerical calculations. Our result is an important milestone towards artificial topological superconducting systems, where Kitaev-like chains\cite{sau2012realizing,kitaev2001unpaired} are assembled from sites hybridized via SCs\cite{leijnse2012parity,su2017andreev,vaitiekenas2020flux}. With the strong coupling demonstrated here, double InAs nanowires can be also promising candidates to host non-Abelian excitations, like parafermions\cite{klinovaja2014time} as a key ingredient of topological quantum computation\cite{kitaev2003fault,nayak2008non}.
 
\section*{Methods}

\textbf{Device fabrication.} InAs nanowires were grown by MBE in the wurtzite phase along the $\langle 0001\rangle$ direction catalyzed by Au. The pattern of the pre-defined Au droplets allowed to control the geometrical properties of the proposed double nanowires, including the diameter, distance, and the corresponding alignment of the cross-sections. The 20 nm thick full-shell Al was evaporated at low temperature in-situ, by rotating the substrate, providing epitaxial, oxide-free layers. The evaporation on such a pair of adjacent nanowires resulted in the merging by the Al. Nanowires with $\sim$80 nm diameter and $\sim$4$ \ \mu$m length were deposited on a p-doped Si wafer capped with 290 nm thick SiO\textsubscript{2} layer by using an optical transfer microscope with micromanipulators. The Al shell on a $\sim$250 nm long segment was removed by means of wet chemical etching. A MMA/MAA EL-6 double-layer performed as a masking layer, which was locally exposed by EBL, allowing the MF-321 selective developer to access the Al (60 s). The etching was followed by a careful localization of the wires with high-resolution SEM. Both source-drain and side gate electrodes were installed in a common EBL step. The sample was exposed to RF Ar milling in the evaporator chamber to remove the native Al\textsubscript{2}O\textsubscript{3}. The process was followed by the metallization of Ti/Al (5/95 nm) without breaking the vacuum. 

\textbf{Experiments.} Low-temperature characterization was carried out in a Leiden Cryogenics dry dilution refrigerator with a base temperature of 40 mK. Transport measurements were performed with standard lock-in technique by applying 10 uV AC signal at $113$ Hz on one of the SC electrodes, whereas the differential conductance was recorded via a home-built current amplifier on the other one. DC bias was adjusted by the offset of the amplifier. We note that due to the geometry, the features of both QDs were measured simultaneously in a single measurement, and hence, the sum of two excitation spectra was captured. Out-of-plane magnetic field was realized by an AMI superconducting magnet.

\textbf{Author contributions.} O. K. and I. L. fabricated the devices, O. K., Z. S. and G. F. performed the measurements and did the data analysis. Z. S. built the theoretical model and developed the numerical simulations. T. K. and J. N. grew the nanowires. All authors discussed the results and worked on the manuscript. P. M. and S. C. proposed the device concept and guided the project.  

\textbf{Acknowledgments.} The authors are thankful to EK MFA for providing their facilities for sample fabrication. We thanks D. Olstein, M. Marnauza, A. Vekris and K. Grove-Rasmussen for experimental assistance, E. T\'ov\'ari, A. P\'alyi, A. Virosztek  for discussion, M. G. Beckerne, F. F\"ul\"op, and M. Hajdu for their technical support. This work has received funding Topograph FlagERA, the SuperTop QuantERA network, the FET Open AndQC and from the OTKA FK-123894 grants. P. M. and G. F. acknowledges support from the Bolyai Fellowship. This research was supported by the Ministry of Innovation and Technology and the NKFIH within the Quantum Information National Laboratory of Hungary and by the Quantum Technology National Excellence Program (Project Nr. 2017-1.2.1-NKP-2017-00001), ÚNKP-20-5 New National Excellence Program, and the Carlsberg Foundation and the Danish National Research Foundation.

\bibliographystyle{naturemag}
\bibliography{and_mol_main_kurtossy_bib}

\end{document}


\renewcommand{\theequation}{S\arabic{equation}}
\renewcommand{\figurename}{Supplementary~Figure}
\renewcommand{\tablename}{Supplementary~Table}
\renewcommand{\thesection}{SUPPLEMENTARY~NOTE~\arabic{section}}

\title{Supplementary Notes for Andreev molecule in parallel InAs nanowires}

\author{Oliv\'er K\"urt\"ossy}
\affiliation{Department of Physics, Budapest University of Technology and Economics and Nanoelectronics 'Momentum' Research Group of the Hungarian Academy of Sciences, Budafoki \'ut 8, 1111 Budapest, Hungary}

\author{Zolt\'an Scher\"ubl}
\affiliation{Department of Physics, Budapest University of Technology and Economics and Nanoelectronics 'Momentum' Research Group of the Hungarian Academy of Sciences, Budafoki \'ut 8, 1111 Budapest, Hungary}

\affiliation{Univ. Grenoble Alpes, CEA, Grenoble INP, IRIG, PHELIQS, 38000 Grenoble, France}

\author{Gerg\H o F\"ul\"op}
\affiliation{Department of Physics, Budapest University of Technology and Economics and Nanoelectronics 'Momentum' Research Group of the Hungarian Academy of Sciences, Budafoki \'ut 8, 1111 Budapest, Hungary}

\author{Istv\'an Endre Luk\'acs}
\affiliation{Center for Energy Research, Institute of Technical Physics and Material Science, Konkoly-Thege Mikl\'os \'ut 29-33., H-1121, Budapest, Hungary}

\author{Thomas Kanne}
\affiliation{Center  for  Quantum  Devices,  Niels  Bohr  Institute,University  of  Copenhagen,  2100  Copenhagen,  Denmark}

\author{Jesper Nyg{\aa}rd}
\affiliation{Center  for  Quantum  Devices,  Niels  Bohr  Institute,University  of  Copenhagen,  2100  Copenhagen,  Denmark}

\author{P\'eter Makk}
\email{peter.makk@mail.bme.hu}
\affiliation{Department of Physics, Budapest University of Technology and Economics and Nanoelectronics 'Momentum' Research Group of the Hungarian Academy of Sciences, Budafoki \'ut 8, 1111 Budapest, Hungary}

\author{Szabolcs Csonka}
\email{szabolcs.csonka@mono.eik.bme.hu}
\affiliation{Department of Physics, Budapest University of Technology and Economics and Nanoelectronics 'Momentum' Research Group of the Hungarian Academy of Sciences, Budafoki \'ut 8, 1111 Budapest, Hungary}

\date{\today}
\maketitle

\section{Non-interacting YSR states (device A)}

\begin{figure}[htp]
\includegraphics[width=1\textwidth]{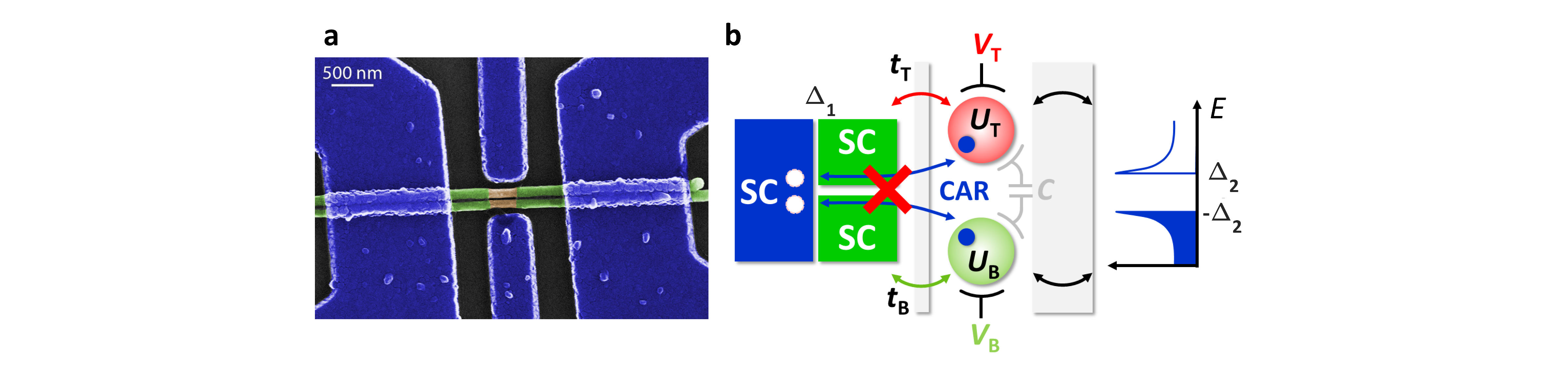}
\caption{\label{supporting_devices} \textbf{Device hosting the non-interacting YSR states.} \textbf{a} False colored SEM micrograph of device A. Compared to device B (see Fig. 1\textbf{a} in the main text), the nanowires originally merged by the eptaxial Al got disconnected and moved further away from each other. \textbf{b} Schematic illustration of device A, similarly to Fig. 1\textbf{d} in the main text. As the nanowires branche (green) are only linked by the common SC (blue), CAR is strongly suppressed by the long distance. Due to the larger spatial separation, the interdot capacitance is also weaker.}
\end{figure}

In the main text, we showed the SEM micrograph and the transport model of the sample exhibiting interacting YSR states, called device B. Here we provide the same for the non-interacting one, device A, in Supp. Fig. \ref{supporting_devices}.

A small gap visible between the wires along the segment covered by epitaxial Al is the indication of the wires being fallen apart during the manipulation process, thus the epitaxial SC link between the wires is missing. On one hand, as the probability of CAR decays with increasing spatial separation of the conducting channels\cite{recher2001andreev}, it is strongly suppressed in device A as it can take place between the QDs only via the ex-situ evaporated common SC. This results in an effective distance of $\sim$800 nm between the QDs. On the other hand, the interdot capacitance of the two InAs branches is also reduced.

\begin{figure}[htp]
\includegraphics[width=1\textwidth]{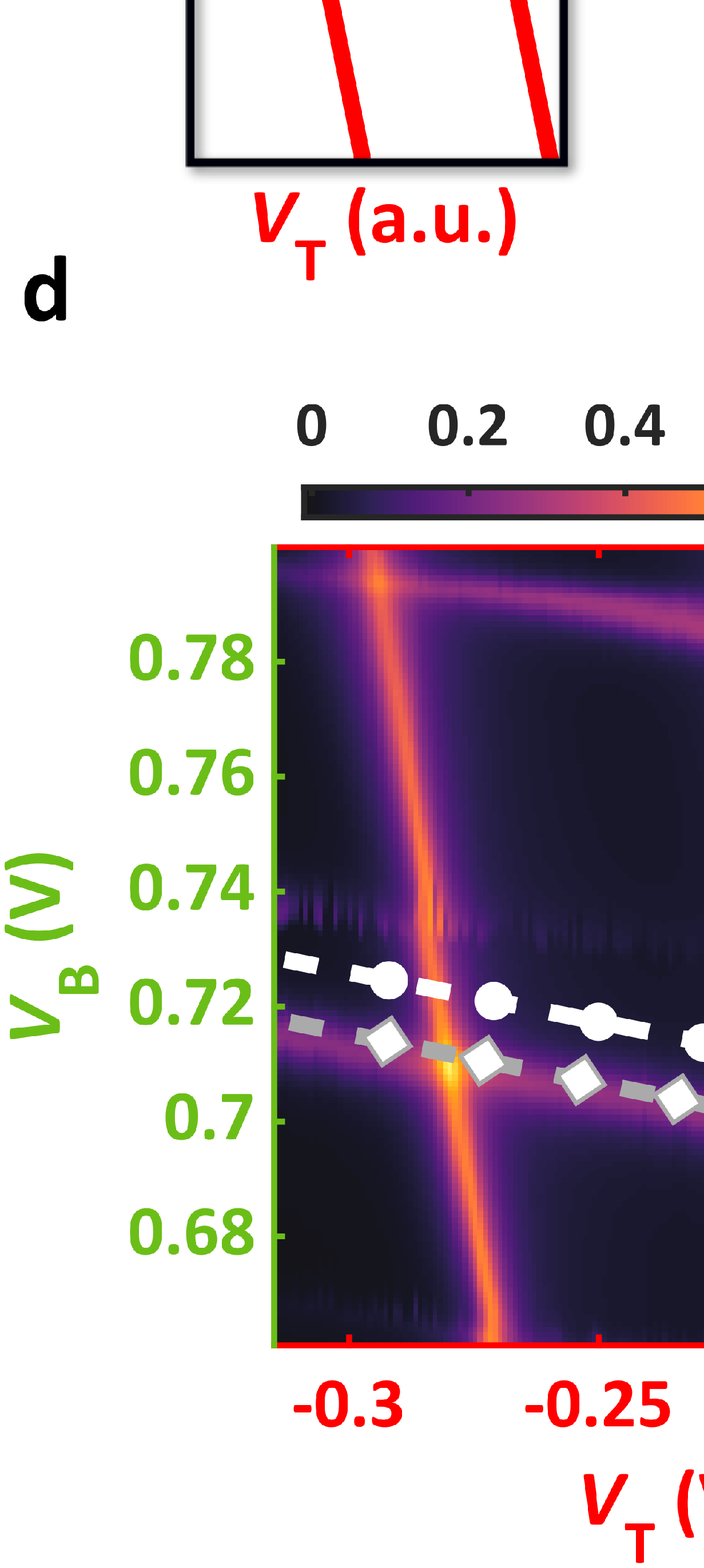}
\caption{\label{supporting_non_int_Shiba} \textbf{Additional data for device A}. \textbf{a} Sketch of the gate stability map as a function of $V_\mathrm{T}$ and $V_\mathrm{B}$. The pink and blue lines indicate the line cuts along the spectra were studied. Non-local bound state spectrum with the markers upon is shown in inset \textbf{I.} to identify the corresponding energies along the cuts. \textbf{b} Predicted excitation spectrum along the pink line in panel \textbf{a}. The local YSR\textsubscript{T} state (red) is tuned by its own plunger gate and develops accordingly, while the non-local YSR\textsubscript{B} state (green) is kept on constant energy due to the parallel slicing with the bottom QD resonances. \textbf{c} Similar spectrum as in panel \textbf{b}, but taken along the blue line, closer to the bottom QD resonance. The reduction of the non-local excitation energy is expected (see green arrows) compared to the one in panel \textbf{b}. \textbf{d} Measured conductance as a function the plunger gates (same map as in Fig. 2\textbf{d}). \textbf{e} Measured excitation spectrum along the white line indicated in panel \textbf{d}. The local and non-local signals (marked by the red and green arrows, respectively) are recovered well and the experiment qualitatively follows the sketch in panel \textbf{b}. \textbf{f} Bias spectroscopy along the gray line in panel \textbf{d}, closer to the resonance of the bottom QD. While the local YSR state (distinguished by the red arrow) is kept intact, the movement of the non-local one is clearly visible as the light green arrow shows its position observed in panel \textbf{e}.}
\end{figure}

The evolution of the YSR states parallel to the top QD resonances was discussed in the main text. Now we demonstrate the expected and measured spectra in device A parallel to the bottom QD resonances. The data is shown in Supp. Fig. \ref{supporting_non_int_Shiba}.

Similarly to Fig. 2, panel \textbf{a} illustrates the conductance as a function of the top and bottom plunger gate voltages in the normal state. Here the spectra are examined parallel to the bottom QD resonances, therefore bias slices along the pink and blue cuts are taken. In this case, the local YSR state resides in the top QD (red, YSR\textsubscript{T}), and the non-local one is attributed to the bottom one (green, YSR\textsubscript{B}), whose energy on the specific slices are also marked in inset \textbf{I.}. We emphasize that YSR\textsubscript{B} is weakly coupled, thus its excitation energy is expected to be at $\Delta_1+\Delta_2$ except in the close vicinity of the resonances where it drops continuously to $\Delta_2$. In panels \textbf{b} and \textbf{c}, the spectra along line cuts are presented off and close to resonance, respectively, with the movement of YSR\textsubscript{B} depicted. Panels \textbf{b}, \textbf{e}, and \textbf{f} show the corresponding transport measurements of the gate stability map, and the bias spectroscopy providing the spectra matching panels \textbf{b} and \textbf{c}, respectively. The local, YSR\textsubscript{T} state adopts (marked by red arrows) the usual "eye shape" and stays unaffected by the movement of the non-local YSR\textsubscript{B} (marked by green arrows) with shifting the slices.

\section{Interacting YSR states (device B)}

\begin{figure}[htp]
\includegraphics[width=0.5\textwidth]{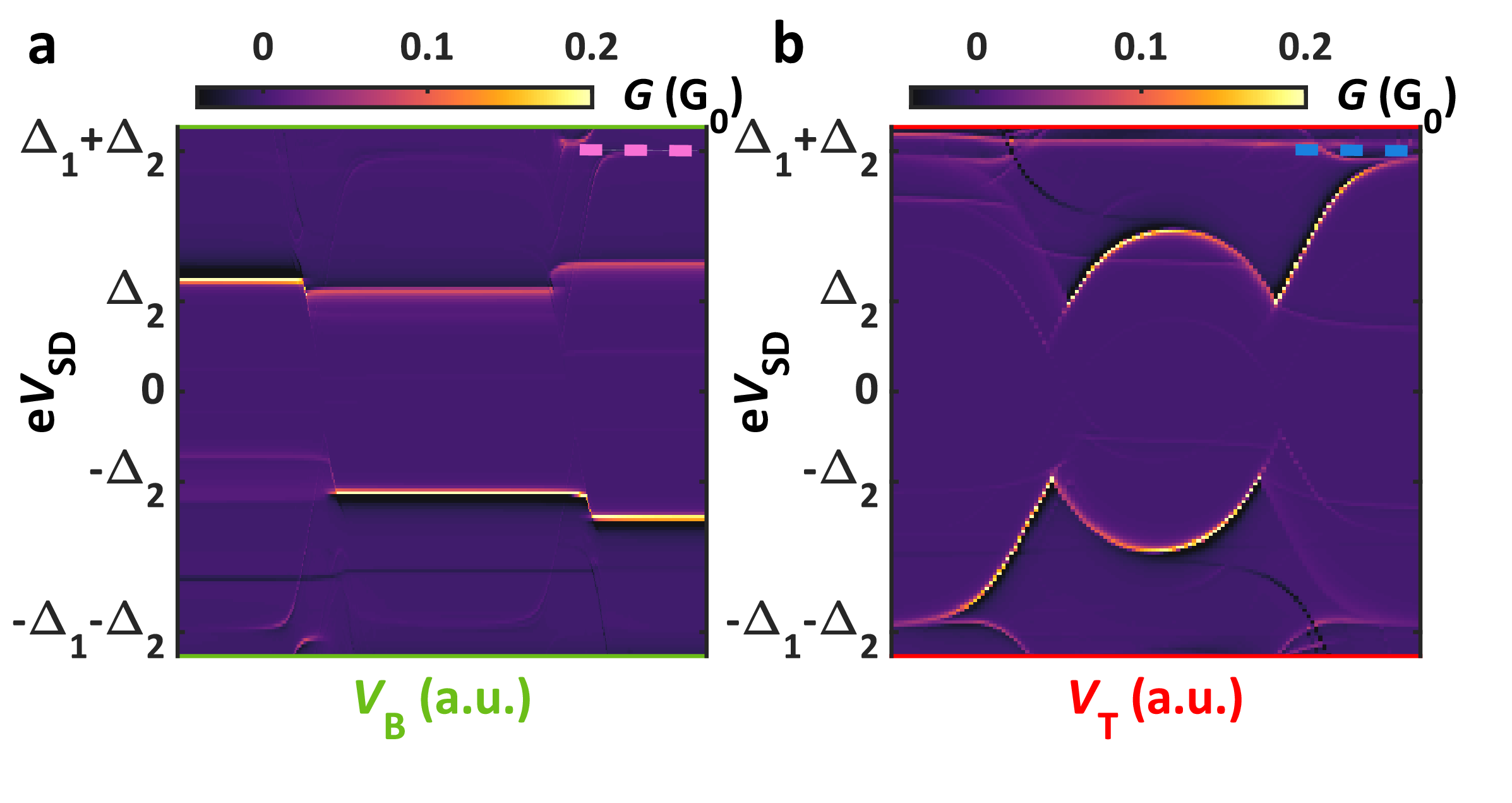}
\caption{\label{supporting_interdot_simulations} \textbf{Simulations of the capacitively interacting YSR states.} \textbf{a} Numeric simulation using the parameter settings of Fig. 3\textbf{c}. The "step-like" shape of the non-local YSR state (YSR\textsubscript{T}) is reproduced. The local YSR state (YSR\textsubscript{B}) is suppressed, as its coupling was chosen three times smaller than the one for YSR\textsubscript{T}. b Similar simulation, but following the gate settings of Fig. 3\textbf{e}. The spectrum of the strongly coupled, local YSR state (YSR\textsubscript{T}) is hardly disturbed by the non-local (YSR\textsubscript{B})
one.}
\end{figure}

In the main text, a pair of bias slices were introduced from the measurements performed on device B revealing the hybridization of the YSR states (see Fig. 3\textbf{e-f}), which were compared to numerical simulations of the fully interacting system introduced in Fig. 4. We present the spectra with the same gate settings with excluding the superconducting coupling and only allowing interdot Coulomb repulsion between the QDs. The results are depicted in Supp. Fig. \ref{supporting_interdot_simulations}.

The simulations in panels \textbf{a} and \textbf{b} are derived along the pink and blue lines in Fig. 3\textbf{a} from the main text. As one can see, the spectra qualitatively match the sketches shown in Fig. 3\textbf{b-c}. Anti-crossings, dispersive lines, and distortions in the local signals observed in the fully interacting case are absent. We note that the excitation lines of YSR\textsubscript{B} are strongly suppressed due to the weak coupling of the bottom QD ($t_\mathrm{T}>t_\mathrm{B}$).

\begin{figure}[htp]
\includegraphics[width=1\textwidth]{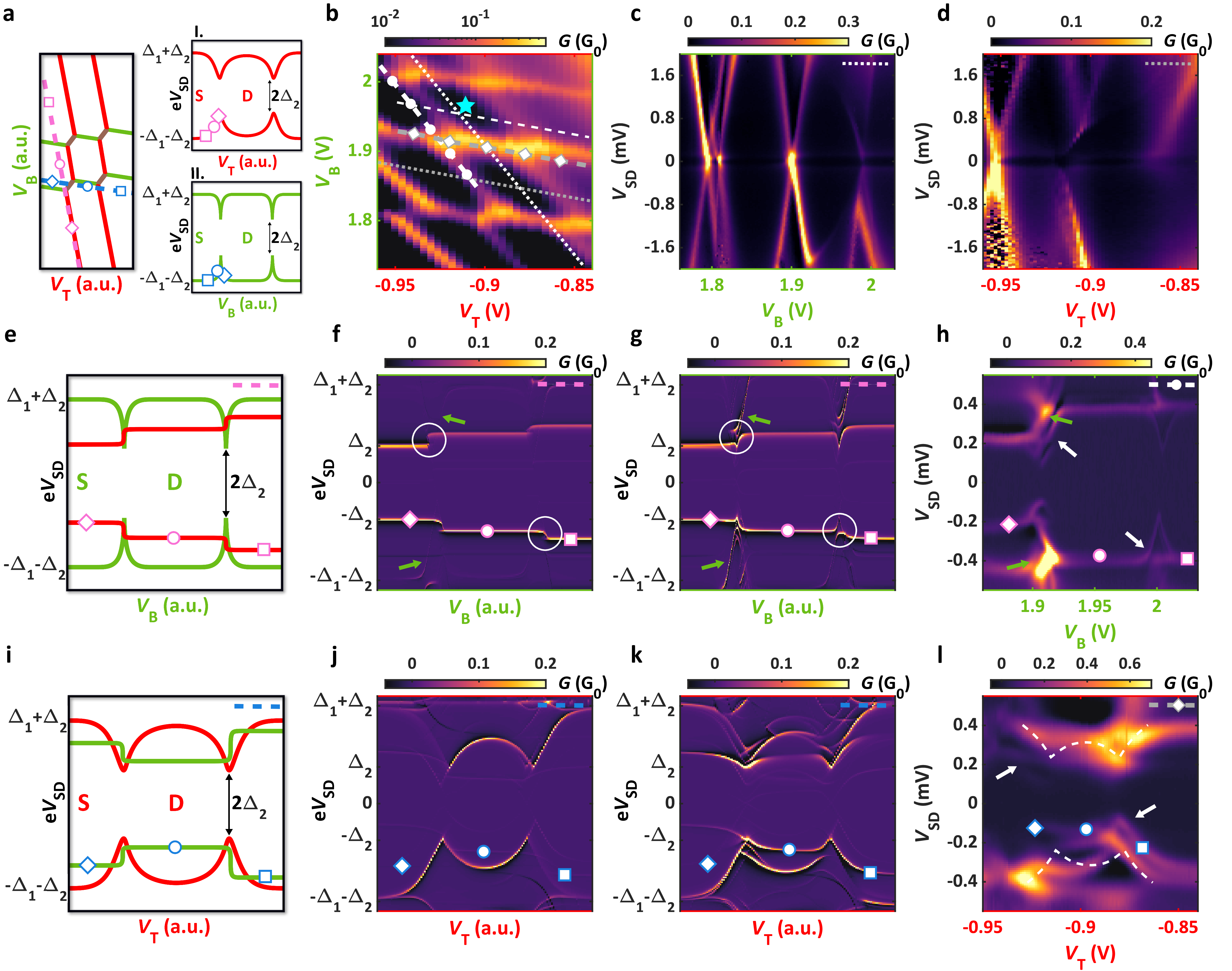}
\caption{\label{supporting_int_Shiba} \textbf{Supplementary data for the interacting YSR states (device B).} \textbf{a} Sketch of the gate stability map as a function of $V_\mathrm{T}$ and $V_\mathrm{B}$ in the presence of strong Coulomb interaction. Similarly to Fig. 3\textbf{a}, the pink and blue lines indicate the line cuts along the spectra were studied. Markers in the inset spectra guide identifying the non-local bound state energies again. \textbf{b} Measured gate stability map in the normal state in a larger window compared to the one in the main text. \textbf{c-d} Large-bias spectroscopy of the QDs in the normal states along the dotted lines in panel \textbf{b}. \textbf{e-f} Predicted and simulated spectra along the pink line in panel \textbf{a} without and \textbf{g} with superconducting coupling introduced in the model. The local, dispersive signal is YSR\textsubscript{B}, the non-local, low-energy one is YSR\textsubscript{T}. The superconducting coupling induces anti-crossings (white circles) and conductance enhancement (green arrows) as seen in the main text data. \textbf{h} Corresponding spectrum captured along the white dashed line in panel \textbf{b}. The highlighted anti-crossings (white arrows) and conductance enhancements (green arrows) are recovered in panel \textbf{g}, in the simulation of the fully interacting model. \textbf{i-k} Similar to panels \textbf{e-g}, but along the blue cut in panel \textbf{a}. The local-signal is YSR\textsubscript{T} providing an "eye-shaped" curve. Completely different development of the non-local signal is expected as the superconducting hybridization is turned on from the purely capacitively interacting case (panel \textbf{j)}. \textbf{l} Finite-bias spectroscopy along the gray dashed line in panel \textbf{b}. The white dashed line shows the undisturbed YSR\textsubscript{T} doublet measured far from the bottom QD resonances. All the key features, including the asymmetry in bias, doubling of the excitation lines and their dispersive evolution are in good agreement with the theory in panel \textbf{k}.}
\end{figure}

Here we provide additional spectra supported by numerical simulations along different traces in the gate stability map. The data is shown in Supp. Fig. \ref{supporting_int_Shiba}.

In panel \textbf{a}, the honey-comb structure as a function of the plunger gate voltages is sketched. Here the pink line is parallel to the top QD resonance. Crossing the bottom QD resonances by increasing $V_\mathrm{B}$ results in an effective gating in the top QD, thus the pink line moves further away from the top QD resonance. The blue line set parallel to the bottom QD resonances follows similar behavior. Inset \textbf{I.} depicts YSR\textsubscript{T} as a function of its own plunger gate, $V_\mathrm{T}$ with the pink square, circle, and diamond markers indicating the current energy along the pink line in the gate stability diagram. Analogously, inset \textbf{II.} depicts YSR\textsubscript{B} as a function of $V_\mathrm{B}$. As shown by the blue diamond, YSR\textsubscript{B} is in the doublet ground state at small $V_\mathrm{T}$. By increasing $V_\mathrm{T}$ the blue line crosses a triple point and YSR\textsubscript{B} ends up in the singlet ground state as indicated by the blue circle and square. Panel \textbf{b} consists of the measured gate stability map in the normal state in a larger window compared to the one in the main text. Panels \textbf{c} and \textbf{d} show large bias spectroscopy measurements accomplished in the normal state along the dotted lines in panel \textbf{b}. Coulomb diamonds were exhibited in a wide gate range in both nanowires and the charging energies were extracted as $U_\mathrm{T}=1.2\,$meV and $U_\mathrm{B}=2.2\,$meV. 

In panels \textbf{e} and \textbf{f} the expected and numerically simulated spectra of the capacitively interacting bound states along the pink cut from panel \textbf{a} are illustrated. Charging energies estimated from the measurements and tunnel amplitudes of $t_\mathrm{T}=0.15\,$meV and $t_\mathrm{B}=0.05\,$meV with $\Delta_1 = 200 \ \mu$V and $\Delta_2 = 120 \ \mu$V were used in the model. The off-site repulsion energy was set to $U_\mathrm{C}=0.1\,$meV. As the distance between the trace and the top QD resonance increases with $V_\mathrm{B}$ (due to effective gating of the Coulomb interaction) in panel \textbf{a}, YSR\textsubscript{T}, which is considered as the non-local signal, jumps from $\sim\Delta_2$ to higher energies when the bottom QD resonances are crossed (see the pink circle and square). Nonetheless, the spectrum changes drastically if superconducting coupling between the QDs is involved (see panel \textbf{g}). Anti-crossings between the local and non-local signals are induced (see white circles), and thus, the latter one bends to lower energy in the vicinity of the charge degeneracy points, similarly to Fig. 4 in the main text. The conductance of the weakly coupled YSR\textsubscript{B} state (marked by the green arrows) is also enhanced in panel \textbf{g} compared to panel \textbf{f}. The corresponding experimental data (panel \textbf{h}) is in much better agreement with the spectrum derived from the fully interacting model. (i) Besides the anti-crossings being dominant (indicated by the white arrows), (ii) the conductance is greatly enhanced at $V_\mathrm{B}=1.91\,$V  predicted by the numerical simulation in panel \textbf{g}. 

The hybridization is even more manifest in the slice parallel to the bottom QD resonances, where the local signal comes from the top QD (YSR\textsubscript{T}), and the non-local does from the bottom one (YSR\textsubscript{B}). Panel \textbf{i} in Supp. Fig. \ref{supporting_int_Shiba} illustrates the naive expectation of the capacitively interacting spectra along the blue cut in panel \textbf{a}. The numerical simulation in panel \textbf{j} qualitatively agrees with the sketch, nevertheless, we note that the shifts in the non-local YSR excitation energies at the ground state transitions are smooth and continuous, and not abrupt. This phenomenon can be attributed to the number of electrons not being quantized in the singlet state since the ground state is the superposition of the empty and double occupied states in a standard YSR or Andreev picture. Panels \textbf{k} and \textbf{l} show the simulation of the fully interacting model and the relevant bias spectroscopy measurement along the gray dashed line in panel \textbf{b}, respectively. In resemblance to the data shown in the main text, (iii) bias asymmetry and (iv) the doubling of the excitation lines with their dispersive evolution (see white arrows in panel \textbf{k}) are observed matching  the theory well in panel \textbf{j}.

\begin{figure}[htp]
\includegraphics[width=1\textwidth]{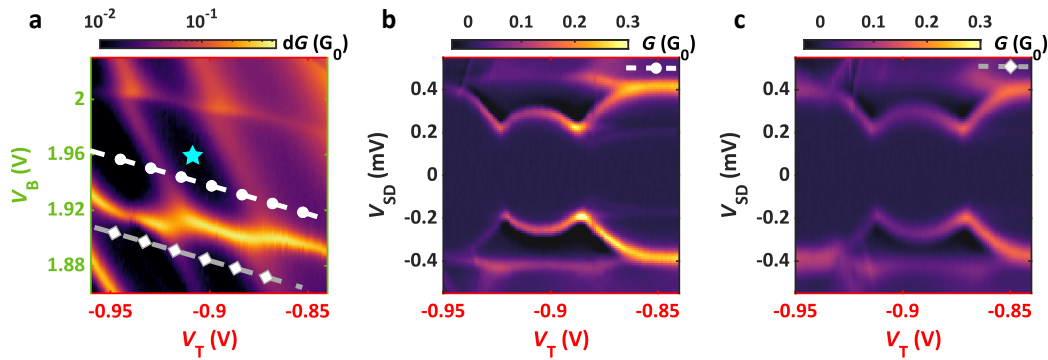}
\caption{\label{supporting_eye_shape} \textbf{Undisturbed YSR state in device B}. \textbf{a} Gate stability map in the normal state (same as the one in Fig. 3 in the main text). The light blue star marks the charges state in \ref{supporting_int_Shiba}\textbf{b}. \textbf{b} Bias spectroscopy measurement along the dashed line from panel \textbf{a}. The local YSR\textsubscript{T} state is unaffected by the non-local YSR\textsubscript{B} one, which is bound to gap edge at $\Delta_1+\Delta_2$ energy. \textbf{c} Similar to panel \textbf{b}, but measured along the other cut from panel \textbf{a} where YSR\textsubscript{B} occupies different ground state parity.}
\end{figure}

We examined the excitation spectra far from any of the bottom QD resonances. Supp. Fig. \ref{supporting_eye_shape}\textbf{a} shows the gate stability map and the traces of the spectroscopy recorded (light blue star identifies the charge state in Supp. Fig. \ref{supporting_int_Shiba}\textbf{b}). As one can see, the lines are selected parallel to the bottom QD resonance, however, they are captured deep in the blockade along the entire map. This resulted in the measured spectra given in panels \textbf{b} and \textbf{c}, where the local "eye-shaped" YSR\textsubscript{T} state is observed without any non-local signals or signatures of hybridization. These excitation lines of the undisturbed YSR\textsubscript{T} doublet are indicated in Fig. 3\textbf{f} and Supp. Fig. \ref{supporting_int_Shiba}\textbf{l} with the white dashed lines. It is also notable that the evolution of the local YSR\textsubscript{T} state is insensitive to the ground state of the non-local, YSR\textsubscript{B} state as it is weakly coupled and evolves regardless of the parity of the electron number. 

The deviation of the YSR states is the strongest when the excitation energies of both YSR states are similar. However, in these particular measurements the bottom QD is in blockade, therefore YSR\textsubscript{B} is bound to the gap edge with the energy of $\Delta_1+\Delta_2$. Consequently, YSR\textsubscript{B} is screened by the quasi-particle continuum, and interaction is suppressed and undetectable\cite{pillet2010andreev}. Therefore the visibility of the hybridization is especially restricted in the spectroscopy measurements where the non-local signal is the weakly coupled YSR\textsubscript{B}.

\section{Andreev molecule in the Kondo regime}

\begin{figure}[htp]
\includegraphics[width=1\textwidth]{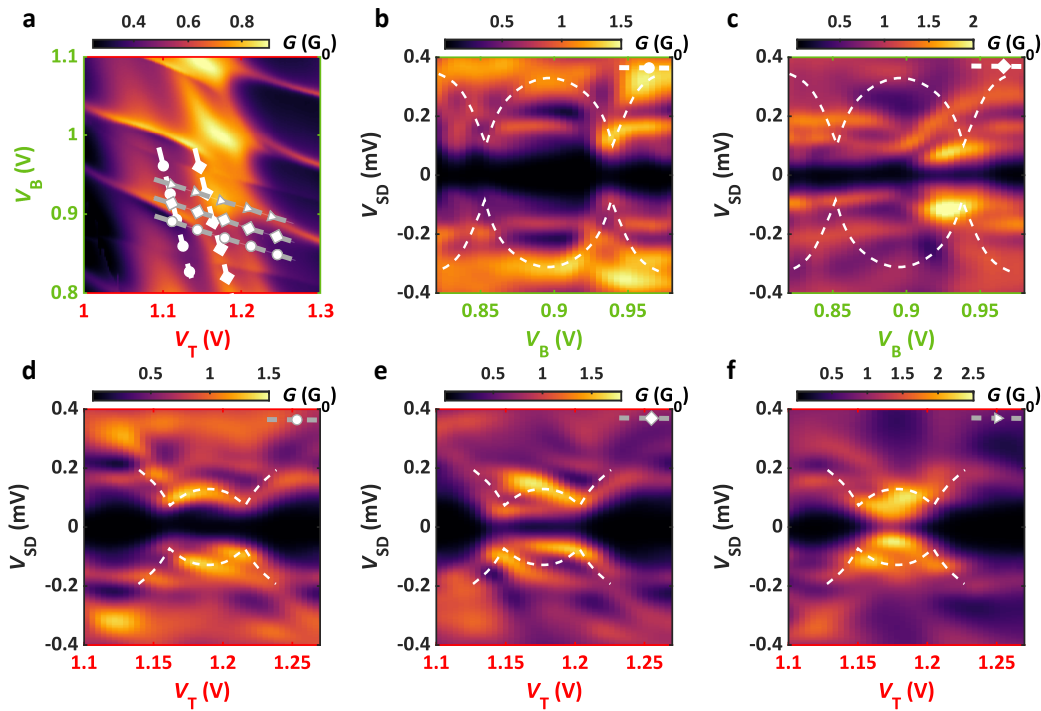}
\caption{\label{supporting_other_gate} \textbf{Interacting YSR states in a different gate range}. \textbf{a} Gate stability sweep in the normal state at more positive plunger gate voltages. The Kondo effect emerged in the top QD. \textbf{b-c} Bias spectroscopy measurements along the white lines in panel \textbf{a}. The local YSR (outlined in panel \textbf{b} with the dashed line) disappeared when the trace of the line cut entered the Kondo regime of the top QD. However, the bendings towards zero energy in the non-local signals were still observable. \textbf{d-f} Measured spectra along the gray lines in panel \textbf{a}. A usual YSR state is distinguishable from the rest in panel \textbf{d} (indicated the white dashed line), which is distorted as the bottom QD is brought to resonance and the conductance is enhanced.}
\end{figure}

We also investigated the bound states in other gate ranges and we observed hybridization between the resonances of the two QDs. Here we report an other interesting example when an Andreev molecule forms in the Kondo regime. Supp. Fig. \ref{supporting_other_gate} summarizes the measurements, which were carried out for more open QDs by applying higher gate voltages. Panel \textbf{a} shows the normal state gate stability map. On one hand, the lack of the blockade at $V_\mathrm{T}=1.2\,$V suggests the top QD being in the Kondo regime, which competes with the superconductivity\cite{van2000kondo, eichler2007even, eichler2009tuning}. On the other hand, further sub-gap states appear in the spectra, therefore the single YSR state picture can not be applied anymore to describe the system. In general, the spectra became more complex due to the multiple excitation lines and their broadenings dropping the visibility of the hybridization. However, signatures of the Andreev molecular state are still present. 

Panels \textbf{b} and \textbf{c} show bias spectroscopy measurements along the white dashed lines in panel \textbf{a}. The line cuts are parallel to the top QD resonances, hence the bottom YSR states are expected to be tuned. While the development of a single YSR state can be followed with the white dashed line in panel \textbf{b}, it does not fit in panel \textbf{c}. Nevertheless, the bendings of low-energy excitations belonging to the top QD ($V_\mathrm{B}=0.92$ V) are still observable.

Panels \textbf{d}, \textbf{e}, and \textbf{f} present the bias spectroscopy measurements along the gray lines in panel \textbf{a}. Now the slicing is parallel to the bottom QD resonances providing the evolution of the top YSR states. A dominant "eye-shaped" excitation appears in the rich spectrum (see the white dashed line in panel \textbf{d}), which changes completely as we approach the bottom QD resonances. While the excitation lines are doubled in panel \textbf{e} as seen earlier in Supp. Fig. \ref{supporting_int_Shiba}\textbf{l}, the local signal deviates from the usual shape and turns into a concave curve from convex one as a function of $V_\mathrm{T}$ in panel \textbf{f} suggesting the strong interaction of the two QDs. 

\newpage

\section{Modeling}

In this Supplementary Note we outline the framework used in the main text to simulate the transport spectrum of the Andreev molecule. First, the Hamiltonian is introduced, then we discuss the transport calculation.

The system consists of two, parallel-coupled quantum QDs and two superconducting electrodes as depicted in Fig. 1\textbf{e}. One of the SC is strongly coupled to the QDs, their hybridization forms the YSR states. The other SC is weakly coupled.

The total Hamiltonian of the system is
\bnen \label{eq:Htot} H = H_{\text{DQD}} + H_{\text{SC1}} + H_{\text{SC2}} + H_{\text{T1}} + H_{\text{T2}}. \eden
The first term describes the double QD,
\bnen \label{eq:HDQD} H_{\text{DQD}} = \sum_{\alpha=\text{T,B}} \left( \varepsilon_{\alpha} n_{\alpha} + U_{\alpha} n_{\alpha\uparrow} n_{\alpha\downarrow} \right) + C n_{\text{T}} n_{\text{B}}, \eden
where $n_{\alpha\sigma} = d^{\dagger}_{\alpha\sigma} d_{\alpha\sigma}$ is the number of electrons with spin $\sigma$ on QD$_{\alpha}$, with $d^{(\dagger)}_{\alpha\sigma}$ being the annihilation (creation) operator of electrons with spin $\sigma$ on QD$_{\alpha}$ and $\alpha=\text{T,B}$ denotes the top, bottom QD. The parameters, $\varepsilon_{\alpha}$ and $U_{\alpha}$ are the level position and the onsite Coulomb energy of QD$_{\alpha}$, and $C$ is the interdot Coulomb repulsion.

The strongly coupled SC is described on the level of the zero bandwidth approximation (ZBA), i.e. it is considered as a single-site SC \cite{JellinggaardPRB2016,AffleckPRB2000,ProbstPRB2016,GroveRasmussenNatComm2018}. This approximation allows for the exact diagonalization of the QD$_{\text{T}}$--SC1--QD$_{\text{B}}$ subsystem's Hamiltonian to obtain the energy spectrum of the Andreev molecule. The ZBA superconductor Hamiltonian is
\bnen \label{eq:HSC1_1} H_{\text{SC1}} = \Delta_1 \left( c^{\dagger}_{\text{SC1}\uparrow} c^{\dagger}_{\text{SC1}\downarrow} + c_{\text{SC1}\downarrow} c_{\text{SC1}\uparrow} \right), \eden
where $c^{(\dagger)}_{\text{SC1}\sigma}$ is the annihilation (creation) operator of an electron with spin $\sigma$ in the SC and $\Delta_1$ is the superconducting gap.
This Hamiltonian can be diagonalized by a Bogoljubov-transformation, $c_{\sigma}=\frac{1}{\sqrt{2}}\left( \gamma_{\sigma} - \sigma \gamma^{\dagger}_{\bar\sigma} \right)$ obtaining
\bnen H_{\text{SC1}} = \Delta_1 \sum_{\sigma} \gamma^{\dagger}_{\sigma} \gamma_{\sigma}. \eden

The tunnel coupling between superconductor SC1 and the QDs writes as
\bnen \label{eq:HT1_1} H_{\text{T1}} = \sum_{\alpha=\text{T,B}} t_{\alpha} \sum_{\sigma} \left( d^{\dagger}_{\alpha\sigma} c_{\text{SC1}\sigma} + c^{\dagger}_{\text{SC1}\sigma} d_{\alpha\sigma} \right), \eden
where $t_{\alpha}$ is tunneling amplitude. Using the Bogoljubov-transformation above this Hamiltonian translates to
 \bnen H_{\text{T1}} = \frac{1}{\sqrt{2}}\sum_{\alpha\sigma} t_{\alpha} \left[ d^{\dagger}_{\alpha\sigma} \left( \gamma_{\sigma} - \sigma \gamma^{\dagger}_{\bar\sigma} \right) + \left( \gamma^{\dagger}_{\sigma} - \sigma \gamma_{\bar\sigma} \right) d_{\alpha\sigma} \right]. \eden

The already defined three Hamiltonian terms, $H_{\text{DQD}} + H_{\text{SC1}} + H_{\text{T1}}$ are numerically diagonalized to obtain the energy spectrum and wavefunction of the Andreev molecular state.

In Eq.~\ref{eq:Htot} $H_{\text{SC2}}$ and $H_{\text{T2}}$ describes the second superconducting lead and its weak tunnel coupling to the QDs, respectively. The superconductor SC2 is described by the BCS Hamiltonian,
\bnen \label{eq:HSC2} H_{\text{SC2}} = \sum_{\vec{k}\sigma} \varepsilon_{\text{SC2}\vec{k}} c^{\dagger}_{\text{SC2}\vec{k}\sigma} c_{\text{SC2}\vec{k}\sigma} + \Delta_2 \sum_{\vec{k}} \left( c^{\dagger}_{\text{SC2}\vec{k}\uparrow} c^{\dagger}_{\text{SC2}-\vec{k}\downarrow} + c_{\text{SC2}-\vec{k}\downarrow} c_{\text{SC2}\vec{k}\uparrow} \right), \eden
where $c^{(\dagger)}_{\text{SC2}\vec{k}\sigma}$ is the annihilation (creation) operator for electrons with momentum $\vec{k}$ and spin $\sigma$ in the SC2 superconductor, $\varepsilon_{\text{SC2}\vec{k}}$ is normal state dispersion and $\Delta_2$ is superconducting gap. Note that here both superconducting gaps, $\Delta_1$ and $\Delta_2$ are assumed to be real, the possible effects originating from the superconducting phase difference are neglected. The tunnel coupling Hamiltonian is
\bnen \label{eq:HT2} H_{\text{T2}} = t_{\text{SC2}} \sum_{\alpha\vec{k}\sigma} \left( d^{\dagger}_{\alpha\sigma} c_{\text{SC2}\vec{k}\sigma} + c^{\dagger}_{\text{SC2}\vec{k}\sigma} d_{\alpha\sigma} \right), \eden
where $t_{\text{SC2}}$ is tunneling amplitude, for the simplicity, assumed to be assumed to be the same for the two QDs.

The tunnel coupling to SC2 is assumed to be weak and treated perturbatively using Fermi's golden rule. In this description the tunnel coupling induces transitions between eigenstates of the QD$_{\text{T}}$--SC1--QD$_{\text{B}}$ system. The time evolution of the occupation of the eigenstates $\ket{\chi}$, $P_{\chi}$ is governed by a master equation together with the normalization condition $\sum_{\chi}P_{\chi}=1$,
\bnen \label{eq:MEq} \frac{dP_{\chi}}{dt} = \sum_{\chi'\neq\chi} \left( W_{\chi\chi'} P_{\chi'} - W_{\chi'\chi} P_{\chi}\right). \eden
Here $W_{\chi\chi'}$ the total transition rate from $\ket{\chi'}$ state to $\ket{\chi}$ induced by the tunnel coupling to SC2. The rates are the sum of two processes, when an electron is added to the Andreev molecule and when one is removed, i.e. $W_{\chi\chi'} = W_{\chi'\chi}\left(d^{\dagger}_{\alpha\sigma} \right) + W_{\chi'\chi}\left(d_{\alpha\sigma} \right) $. These two contributions are expressed as
\bean W_{\chi'\chi}\left(d^{\dagger}_{\alpha\sigma} \right) = \pi t^2_{\text{SC2}} \left|\bra{\chi'} d^{\dagger}_{\alpha\sigma} \ket{\chi} \right|^2 \rho_S\left(E_{\chi}-E_{\chi'}-\mu_{\text{SC2}} \right) f\left(E_{\chi}-E_{\chi'}-\mu_{\text{SC2}} \right) \nonumber \\ 
W_{\chi'\chi}\left(d_{\alpha\sigma} \right) = \pi t^2_{\text{SC2}} \left|\bra{\chi'} d_{\alpha\sigma} \ket{\chi} \right|^2 \rho_S\left(E_{\chi}-E_{\chi'}+\mu_{\text{SC2}} \right) f\left(E_{\chi}-E_{\chi'}+\mu_{\text{SC2}} \right), \eean
where $\rho_{S}(E) = \rho_0 \text{Re}\frac{E+i\gamma}{\sqrt{\left(E+i\gamma\right)^2 - \Delta^2_2}}$ is the Dyson-like density of states (DOS) in SC2 superconductor, with $\gamma$ being the Dyson-parameter and $\rho_0$ is the normal state DOS, assumed to be constant, $f(E)$ is the Fermi function, $E_{\chi}$ denote the energy of the $\ket{\chi}$ state and $\mu_{\text{SC2}}=eV_{\text{SD}}$ is chemical potential difference of the two superconducting leads, due to the applied bias voltage, $V_{\text{SD}}$.

To derive the current through the device the master equation, Eq.~\ref{eq:MEq} is solved in the stationary limit, $dP_{\chi}/dt=0$ to obtain the occupations. The current writes as
\bnen I = \frac{e}{\hbar} \sum_{\alpha\chi\chi'\sigma} \left[ W_{\chi'\chi}\left(d_{\alpha\sigma} \right) - W_{\chi'\chi}\left(d^{\dagger}_{\alpha\sigma} \right) \right] P_{\chi}. \eden
The differential conductance is obtained as the derivative of the current, i.e. $G = e\frac{dI}{d\mu_{\text{SC2}}}$.

For the case, when only capacitive coupling is assumed between the QDs, one has to eliminate the SC mediated tunneling processes that couples the states of the two QDs. An example for such a process is when a Cooper pair from the SC1 electrode splits between the QDs. As the QD$_{\text{T}}$--SC1--QD$_{\text{B}}$ subsystem is treated up to all orders in the tunneling, such processes are necessarily present in the description above. One can formally remove them by coupling the QDs to two separate SCs. This can be formulated in the following Hamiltonians:
\bnen H_{\text{SC1}} = \Delta_1 \sum_{\alpha=\text{T,B}} \left( c^{\dagger}_{\text{SC1}\alpha\uparrow} c^{\dagger}_{\text{SC1}\alpha\downarrow} + c_{\text{SC1}\alpha\downarrow} c_{\text{SC1}\alpha\uparrow} \right), \eden
\bnen H_{\text{T1}} = \sum_{\alpha=u,l} t_{\text{SC1}\alpha} \sum_{\sigma} \left( d^{\dagger}_{\alpha\sigma} c_{SC1\alpha\sigma} + c^{\dagger}_{SC1\alpha\sigma} d_{\alpha\sigma} \right). \eden
The difference compared to Eqs.~\ref{eq:HSC1_1}\&\ref{eq:HT1_1} is that the SC1 superconductor is split into two parts, which are only coupled to one of the QDs, This way the tunnel coupling hybridize the QD and SC states to local YSR states, but the further hybridization of the YSR state into molecular states are prevented. The rest of the transport calculation is the same as above. 

\bibliographystyle{naturemag}
\bibliography{and_mol_supp_kurtossy_bib}

